\documentstyle[12pt]{article}
\def\smal{}
\def\sss{}
\def\Comm#1{{\tt{[#1]}}}

\def\ep{\epsilon}

\def\a{{\bf w}}

\def\q{\quad}

\def\ie{{\it i.e.}}
\def\eg{{\it e.g.}}
\catcode`@=11
\newdimen\z@ \z@=0pt
\def\m@th{\mathsurround=\z@}
\def\ialign{\everycr{}\tabskip\z@skip\halign} 
\def\eqalign#1{\null\,\vcenter{\openup\jot\m@th
  \ialign{\strut\hfil$\displaystyle{##}$&$\displaystyle{{}##}$\hfil
      \crcr#1\crcr}}\,}
\def\matrix#1{\null\,\vcenter{\normalbaselines\m@th
    \ialign{\hfil$##$\hfil&&\quad\hfil$##$\hfil\crcr
      \mathstrut\crcr\noalign{\kern-\baselineskip}
      #1\crcr\mathstrut\crcr\noalign{\kern-\baselineskip}}}\,}
\catcode`@=12
\def\be{\begin{equation}}
\def\ee{\end{equation}}
\def\eq#1{{(\ref{#1})}}

\begin{document}
\textheight 9in
\topmargin 0in
\begin{flushright}
DAMTP/97-19
\vskip0pt hep-th/9703137

\end{flushright} 
\vskip20pt
\begin{center}
\begin{Large}
{\bf Intrinsic anyonic spin through deformed geometry}\vskip0pt
\end{Large}
\vspace{1cm}
{\bf R.S. Dunne\footnote{e-mail: r.s.dunne@damtp.cam.ac.uk}}
\vskip10pt
{\it Department of Applied Mathematics \& Theoretical Physics}\vskip0pt
{\it University of Cambridge, Cambridge CB3 9EW}
\end{center}

\begin{abstract}
The properties of the deformed bosonic oscillator, and the quantum groups ${\cal U}_q(SL(2))$
and $GL_q(2)$ in the limit as their deformation parameter $q$ goes to
a root of unity are investigated and interpreted physically.
These properties are seen to be related to fractional supersymmetry and intrinsic
anyonic spin. A simple deformation of the Klein-Gordon equation is
introduced, based on $GL_q(2)$. When $q$ is a root of unity this
equation is a root of the undeformed Klein-Gordon equation.

\vskip20pt\noindent
\end{abstract}
\def\smal{}
\def\U{{\cal{U}}}
\def\chh{\chi}
\def\ul{\underline}
\def\omw{w}
\def\dual{{\cal K}}
\def\[{[\![}
\def\]{]\!]}
\def\li{q\to-1}
\def\ff{f(\theta)}
\def\drr{{d\over d\theta}}
\def\ep {\parindent=0pt\vfill\eject}
\def\np {\parindent=20pt}
\def\s {\vskip0pt\noindent}
\def\bb #1 {\vskip0pt\noindent[#1]\hskip5pt}
\def\r #1 {\hskip5pt[#1]\hskip3pt}
\def\exp{{\rm exp}}
\def\ii{{\rm id}}
\def\a{{a^\dagger}}
\def\b{{b^\dagger}}
\def\tta{{\theta^{n-1}\over[n-1]_q!}}
\def\sn{{\sum_{m=0}^{n-1}}}
\def\sr{{\sum_{m=0}^{r}}}
\def\sno{{\sum_{m=0}^{n-1}}}
\def\pt{{\partial\over\partial\theta}}
\def\dt{{\partial\over \partial t}}
\def\dr{{d_R\over d_R\theta}}
\def\dl{{d_L\over d_L\theta}}
\def\dz{\partial_t}
\def\qq{\exp({2\pi i\over n})}
\def\qm{\lim_{q\to-1}}
\def\ql{\lim_{q\to\qq}}
\def\qmm{{\lim\over{q\to-1}}}
\def\ad{a^\dagger}
\def\ff{f(\theta)}
\def\zz{\vert0\rangle}
\def\rr{{d_R\over d_R\theta}}
\def\ll{{d_L\over d_L \theta}}
\def\gg{{q^{g(A_m)}}}
\def\ssn{{\sum_{n=0}^\infty}}
\def\ss{{\sum_{m=0}^\infty}}
\def\d{{{\cal D}_L}}
\def\dR{{{\cal D}_R}}
%
\def\sss{}
\def\q{\quad}
\def\ie{{\it i.e.}}
\def\eg{{\it e.g.}}
\catcode`@=11
\newdimen\z@ \z@=0pt
\def\m@th{\mathsurround=\z@}
\def\ialign{\everycr{}\tabskip\z@skip\halign} 
\def\eqalign#1{\null\,\vcenter{\openup\jot\m@th
  \ialign{\strut\hfil$\displaystyle{##}$&$\displaystyle{{}##}$\hfil
      \crcr#1\crcr}}\,}
\def\matrix#1{\null\,\vcenter{\normalbaselines\m@th
    \ialign{\hfil$##$\hfil&&\quad\hfil$##$\hfil\crcr
      \mathstrut\crcr\noalign{\kern-\baselineskip}
      #1\crcr\mathstrut\crcr\noalign{\kern-\baselineskip}}}\,}
\catcode`@=12
\def\be{\begin{equation}}
\def\ee{\end{equation}}
\def\eq#1{{(\ref{#1})}}
\def\Comm#1{{\tt{[#1]}}}
\def\ie{{\it i.e.}}
\def\eg{{\it e.g.}}

\def\U{{\cal{U}}}
\def\am{{a_{-}}}
\def\ap{{a_{+}}}
\def\Am{{A_-}}
\def\Ap{{A_+}}
\def\NA{N_A}
\def\bm{{b_-}}
\def\bp{{b_+}}
\def\ql{$q\to\epsilon$ }
\def\qe{$q=\epsilon$ }
\def\qlm{\lim_{q\to\epsilon}}
\def\zz{\vert0\rangle}
\def\zzr{\langle0\vert}
\def\qo{$q$-oscillator}
\def\ad{(a^\dagger)}
\def\bd{(b^\dagger)}
\def\Ad{(A^\dagger)}

\def\Comm#1{{\tt{[#1]}}}
\def\alp{\alpha}
\def\bet{\beta}
\def\ifff{\rm for}
\def\*{{$*$}}
\def\det{{\rm det}}
\def\sp{J_{+}}
\def\sm{J_{-}}
\def\spm{J_{\pm}}
\def\sz{J_{z}}

\def\lp{L_{+}}
\def\lm{L_{-}}
\def\lpm{L_{\pm}}
\def\lz{L_{z}}

\def\n{n'}
\def\aom{a_{1-}}
\def\atm{a_{2-}}
\def\aop{a_{1+}}
\def\atp{a_{2+}}

\def\bom{b_{1-}}
\def\btm{b_{2-}}
\def\bop{b_{1+}}
\def\btp{b_{2+}}

\def\akp{a_{k+}}
\def\akm{a_{k-}}

\def\mod{{\rm mod}}
\def\divv{{\rm div}}

\def\q{\quad}

\def\A{{\cal A}}
\def\B{{\cal B}}
\def\C{{\cal C}}
\def\D{{\cal D}}
\def\]{]\!]}
\def\sym{\alpha}

\sss
\vfill\eject
\section{Introduction}
In a series of recent papers \cite{DMAP1,DMAP2,DMAP3,DMAP4,RSD}, a novel interpretation of supersymmetry
and its generalization to fractional supersymmetry has been developed. The observation
upon which these results are based is that when the deformation parameter $q$ 
is a root of unity, there are circumstances in which twice as many variables as usual are 
needed to fully describe the
braided line \cite{MajI,MajII,MajIII}, the extra variables being essential
if one is to retain in this limit all of the 
structure associated with the braided line at generic $q$ 
(for the  purposes of 
the present paper generic $q$ is taken to mean $\vert
q\vert=1$, 
with $q$ not equal to a root of unity). It was remarked in various places in these papers that this basic feature
is likely to be widespread in the theory of Hopf algebras and braided Hopf algebras,
and the present paper is devoted to working out some new and physically interesting
examples.\par
We begin in section 3 by taking the $q\to\epsilon$ ($\epsilon$ a nontrivial
primitive ${\tilde n}$th root of  unity) limit of the $q$-deformed
bosonic 
oscillator \cite{AC,Mac1,Biedenharn} or $q$-oscillator, and establish
that 
in this limit it decomposes into two independent oscillators, one of
them an ordinary (undeformed) boson and the other an anyon (we use the
word 
anyon in the sense of \cite{MajIII}). The corresponding
decomposition of the Fock space when $q\to\epsilon$ is also studied. 
For certain values of $q$
it turns out that  in this limit both parts of the decomposed Fock space 
(it now has the form of a direct
product) are physical (\ie\ all states have positive definite norm). Since there exist
$q$-oscillator realizations of all of the deformed enveloping algebras $\U_q(g)$
\cite{Hayashi}, it is
reasonable to expect these to exhibit analogous decompositions when $q\to\epsilon$.\par

In section 4 we illustrate this for 
the simplest case, $\U_q(sl(2))$. Utilizing the $q$-Schwinger realization \cite{Mac1}
 we establish, for this realization, and the corresponding 
highest weight  representations the decomposition of
$\U_q(sl(2))$ into the direct product of undeformed $\U(sl(2))$, and $\U_\epsilon(sl(2))$
the naive version of $\U_q(sl(2))$ at $q=\epsilon$ obtained by simply setting
$q=\epsilon$. To go further we draw on the detailed studies of $\U_q(sl(2))$ when $q$ is
a root of unity, carried out by Kac \cite{KAC}, Lusztig \cite{LUSZTIG1,LUSZTIG2}, 
Arnaudon \cite{Arnaudon1}, and collaborators. Using these we place 
restrictions on 
$\U_q(sl(2))$ as a Hopf algebra at $q=\epsilon$, finally obtaining a restricted form of the
$\U_q(sl(2))$ Hopf algebra which has
irreducible representations identical (up to a constant factor) 
to those associated with the Schwinger realization. We then extend
this Hopf algebra, introducing new elements which are endowed with that part of the 
Hopf structure of $\U_q(sl(2))$ which is usually lost when we set $q=\epsilon$, and obtain
thereby an FSUSY-like generalization of the $su(2)$ symmetry algebra. Although
the extended algebra does not have direct product form, all of its highest weight  representations do, and
using our earlier work with the Schwinger realization, we are able to choose the basis
in which this is most clearly manifested. This enables us to interpret
such highest weight  
representations as being the
direct product of an intrinsic spin degree of freedom and an orbital angular momentum degree of
freedom.\par

It is reasonable to expect analogous results for $SL_q(2)$, the Hopf algebra dual to $\U_q(sl(2))$
\cite{FRT}.
In fact the extension to all of $GL_q(2)$ is not difficult, and this is the
topic of section 5. When $q$ is a primitive ${\tilde n}$th root of unity, we find that the
unrestricted quantum group $GL_q(2)$, the matrices of which we denote
by $\omw$, 
has sub-Hopf algebra $GL_{q^{n^2}}(2)$ (here $n={\tilde n}$ for odd
${\tilde n}$ a
nd $n={{\tilde n}\over2}$ for even ${\tilde n}$),
the matrices of which we denote by $W$. In the case of odd ${\tilde n}$ this sub-Hopf algebra is 
identical
to the undeformed quantum group $GL(2)$, and moreover, it is central. Setting det$_q\omw=1$ leads
directly to det$_{q^{n^2}}W=1$, so that all of these results hold equally for $SL_q(2)$.
By imposing a suitable \*-structure on $GL_q(2)$, we find that for $q^{n^2}=1$ we can identify the elements of 
$W$ with linear combinations
of the momenta of undeformed Minkowski spacetime. The quantum determinant of $W$ is just the
undeformed Laplacian, and the quantum determinant of $\omw$ has the form of a $q$-deformed Laplacian.

This deformed Laplacian turns out to be an $n$th root of the undeformed Laplacian, an observation
which enables us to construct (anyonic) Dirac-like square roots of the Klein-Gordon equation.
\par
A restricted version of $GL_q(2)$ can also be built. In this case $W$ only contains two
elements. These are grouplike and, for $SL_q(2)$, mutually inverse. As in the case of
$\U_q(sl(2))$ we can extend the restricted version of $GL_q(2)$ by elements which carry the Hopf
structure lost from the generic case. We thereby obtain a new quantum group which involves a
FSUSY-like coproduct.\par

\section{Notation and conventions}

In this section we establish our notation and
a few results of which we will later have need. 
We begin by defining,
\be
\[X\]={q^X-q^{-X}\over q-q^{-1}}=q^{1-X}[X]_{q^2}\q,                                         \label{ii.II}   
\ee
A consequence of the relationship between $\[X\]$ and $[X]_{q^2}$,
is that the $q$ used in the present paper corresponds to the square root of the $q$ 
used in \cite{DMAP1,DMAP2,DMAP3,DMAP4,RSD}, so that for example super/fermionic 
properties are associated here with $q=\pm i$ rather than with $q=-1$.  
For nonnegative integer r we also define
\be\eqalign{
\[r\]!&=\left\{\eqalign{&\[r\]\[r-1\]\[r-2\]...\[2\]\[1\]\q,\quad
{\rm for\hskip5pt integer\hskip5pt}r>0\q,\cr
&1\q,\quad {\rm for\hskip5pt} r=0\q,\cr
}\right.\cr
\left[\!\!\left[\matrix{m\cr r\cr}\right]\!\!\right]&
={\[m\]!\over\[m-r\]!\[r\]!}\q.}                    \label{ii.III}
\ee
These factorials are related to the $[r]_q!$ factorials given in
\cite{DMAP1,DMAP2} via
\be
\eqalign{
[m]_{q^2}!&=q^{m(m-1)\over2}\[m\]!\q,\cr
[m]_{q^{-2}}!&=q^{-m(m-1)\over2}\[m\]!\q.                                 \label{ai}
}\ee
  Let us define $\epsilon$ to be a primitive ${\tilde n}$th root of unity
so that $\epsilon^{\tilde n}=1$, where ${\tilde n}>2$ is a positive integer. The cases of odd
and even ${\tilde n}$ have to be treated in slightly different ways and because of this it is
useful to introduce a variable $n$, defined by
\be 
n=\left\{\eqalign{&{\tilde n}{\rm\hskip15pt for\hskip5pt 
odd\hskip5pt} {\tilde n}\q,\cr
&{{\tilde n}\over2}{\rm\hskip15pt 
for\hskip5pt even\hskip5pt} {\tilde n}\q,}\right.
                                                                     \label{v.IIsec} 
\ee 
so that for odd ${\tilde n}$, $\epsilon^{n}=1$ and for 
even ${\tilde n}$, $\epsilon^{n}=-1$. For even
$n$ the roots $q=\exp(\pm\frac{i\pi}{n})$ are of 
particular interest since for $0< p< n$ the
quantity $\[p\]$ is positive definite,
\be
\[p\]={\sin{\pi p\over n}\over \sin {\pi\over n}}\q,                             \label{ii.IV}
\ee
and this has important consequences for the construction 
of unitary representations. For similar reasons we
also note that with $q=\exp(\frac{2\pi ia}{{\tilde n}})$ 
the quantity $\[m\]!$ takes negative values for the first time
when $m>\frac{{\tilde n}}{2a}$.
Also, for $r\geq0$ and $n>p\geq0$ we have $\[rn\]$=0, and the following useful identities,
\be\eqalign{
{\[rn+p\]\over \[p\]}
&=q^{rn}\left({1-q^{(2nr+2p)}\over 1-q^{2p}}\right)\q\cr
&=q^{rn},}                                                                 \label{ii.V}
\ee
and 
\be\eqalign{
\qlm {\[rn\]\over\[r\]}
&=\qlm q^{(1-r)n}\left({1-q^{2rn}\over 1-q^{2n}}\right)\q\cr
&=\qlm q^{(1-r)n}(1+q^{2n}+q^{4n}+...+q^{2n(r-1)})\q\cr
&=q^{n(1-r)}r\q.}                                                            \label{ii.VI}
\ee

\section{Deformed bosons at $q$ a root of unity}

The $q$-deformed bosonic oscillator (\qo) has been
 discussed by numerous authors \cite{AC,Mac1,Biedenharn}. 
It is defined by the relations 
\be 
\am\ap-{q^{\pm 1}\ap\am}=q^{\mp{N}}\q,
\hskip20pt [N,a_{\pm}]=\pm a_{\pm}\q.            \label{iii.I}
\ee
It follows immediately from this definition that
\be
[q^N\am,\ap]_{q^2}=1\q,\hskip20pt   
 [q^N\am,(\ap)^m]_{q^{2m}}=[m]_{q^2}(\ap)^{m-1}\q,      \label{iii.VII}
\ee
and hence
\be\eqalign{
[q^N\am,[q^N\am,[\ldots[q^N\am,(\ap)^r]_{q^{2r}}\ldots]_{q^4}]_{q^2}]
&=[r]_{q^2}!\q\cr
&=q^{\frac{r(r-1)}{2}}\[r\]!\q.}                                      \label{iii.VIII}
\ee
We now set $r=n$ and take the \ql limit, obtaining
\be\eqalign{
\qlm\frac{1}{\[n\]!}[q^N\am,[q^N\am,[\ldots[q^N\am,(\ap)^n]_{q^{2n}}\ldots]_{q^4}]_{q^2}]
&=\qlm {q^{\frac{n(n-1)}{2}}\over\[n\]!}[q^{nN}(\am)^n,(\ap)^n]\q\cr
&=q^{\frac{n(n-1)}{2}}\q,}                                            \label{iii.IX}
\ee
which we can write as 
\be
\qlm \left[q^{\frac{nN}{2}}{(\am)^n\over\sqrt{\[n\]!}}\q,
{(\ap)^n\over\sqrt{\[n\]!}}q^{\frac{nN}{2}}\right]=1\q.                       \label{iii.X}
\ee
 Note that since $q^{2n}=1$ we can write $q^{nN}=q^{-nN}$. In consequence it is possible to change the signs on the 
 exponents of the $q^{\frac{nN}{2}}$ terms in the above and in the following definitions, 
a freedom of which we will
shortly make use. Motivated by \eq{iii.X} we now define
\be
\bp=\qlm{(\ap)^n\over \sqrt{\[n\]!}}q^{\frac{nN}{2}}\q,\hskip20pt
\bm=\qlm q^{\frac{nN}{2}}{(\am)^n\over \sqrt{\[n\]!}}\q.                       \label{iii.XI}
\ee
Then from (\ref{iii.X})
\be
[\bm,\bp]=1\q,                                                                     \label{iii.XII}
\ee
which is just the defining relationship of 
an ordinary boson.

If, in the following,  we want 
to work at an algebraic level, we must assume that 
$(\ap)^n\to 0$  and $(\am)^n\to 0$ when \ql in such a
way that $\bp$ and $\bm$ are well defined, or put another
way we must restrict our attention to the subalgebra for which
this is true. We make such a restriction, and note that for 
Fock space representations it follows automatically.  
Let us now calculate the commutation relations between these
new bosonic oscillators and our original $q$-oscillators.
Trivially we have $[\am,\bm]_{q^{\frac{n}{2}}}=0$ and 
$[\ap,\bp]_{q^{-\frac{n}{2}}}=0$.
The other two commutation relations can be obtained as follows,
\be\eqalign{
[\am,\bp]_{q^{-\frac{n}{2}}}
&=\qlm\left[\am,{(\ap)^n\over \sqrt{\[n\]!}}{q^{\frac{nN}{2}}}\right]_{q^{-\frac{n}{2}}}\q,\cr
&=\qlm q^{-N}\left[q^N \am,{(\ap)^n\over \sqrt{\[n\]!}}\right]q^{\frac{nN}{2}}\q,\cr
&=\qlm q^{-N}\sqrt{\[n\]}{(\ap)^{n-1}\over \sqrt{\[n-1\]!}}q^{\frac{nN}{2}}\q,\cr
&=0\q.}                                                                           \label{iii.XIII}
\ee
Here we have made use of (\ref{iii.VII}) in going from 
the second line to the third. Similarly we find
$[\ap,\bm]_{q^{\frac{n}{2}}}=0$, so the complete set of commutation
relations between the two sets of oscillators is as follows,
\be\eqalign{
[\am,\bm]_{q^{\frac{n}{2}}}&=0\q,\hskip20pt
[\am,\bp]_{q^{-\frac{n}{2}}}=0\q,\cr
[\ap,\bm]_{q^{\frac{n}{2}}}&=0\q,\hskip20pt
[\ap,\bp]_{q^{-\frac{n}{2}}}=0\q.}                                                   \label{iii.XIV}
\ee
Introducing a number operator for these new bosonic
oscillators, defined in the usual way as $N_b=\bp\bm$,
we also have
\be\eqalign{
[N,a_\pm]&=\pm a_\pm\q,\hskip20pt
[N_b,a_\pm]=0\q,\cr
[N,b_\pm]&=\pm n b_\pm\q,\hskip20pt 
[N_b,b_\pm]=\pm b_\pm\q.}                                                     \label{iii.XV}
\ee
We can use these results to choose a more natural
set of generators for the algebra. Let us define
\be\eqalign{
\Am&=\am{q^{-\frac{nN_b}{2}}}\q,\cr
\Ap&={q^{-\frac{nN_b}{2}}}\ap\q,}                                                  \label{iii.XVI}
\ee
and
\be
\NA=N-nN_b\q.                                                                     \label{iii.XVII}
\ee
Then in terms
of these new generators we have,
\be\eqalign{
[\Am,\Ap]_{q^{\pm1}}&=q^{\mp\NA}\q,\hskip20pt
[\NA,A_{\pm}]=\pm A_{\pm}\q,\cr
[\bm,\bp]&=1\q,\hskip28pt
[N_b,b_\pm]=\pm b_\pm\q,}                                                    \label{iii.XVIII} 
\ee
which are the defining relations of an anyon and an ordinary undeformed boson, as well as,
\be
 [A_\pm,b_{(\pm)}]=0\q,\quad
[\NA,b_\pm]=0\q,\quad
[N_b,A_\pm]=0\q,\quad
[N_b,\NA]=0\q,                                                            \label{iii.XIX}
\ee
which show that the two algebras commute.
Thus in the $q\to\epsilon$ limit the deformed bosonic oscillator algebra decomposes into the direct product of the undeformed bosonic oscillator algebra and an anyonic oscillator algebra. A clearer understanding
of the way in which the splitting of a single $q$-oscillator
at generic $q$, into two independent oscillators when $q\to\epsilon$
occurs, can be obtained by examining the corresponding result for
the Fock space. At generic $q$, the normalized state $\vert m\rangle$
is defined by
\be
\vert m\rangle={(\ap)^m\over \sqrt{\[m\]_q!}}\vert0\rangle\q,                   \label{iii.XX}
\ee   
where $a\vert0\rangle$=0. If we write $m=rn+p$, 
for integers $0\leq p<n$, $r\geq0$, then after a little algebra
(\ref{iii.XX}) can be written as
\be
\vert rn+p\rangle= (\ap)^p\left({(\ap)^n q^\frac{nN}{2}\over \sqrt{\[n\]_q!}}\right)^r
\left({(\[n\]_q!)^r\over\[rn+p\]_q!}\right)^{1\over2}
\prod_{\alpha=0}^{r-1}q^{-{n^2\alpha\over2}}\vert0\rangle\q.                       \label{iii.XXI}
\ee
Also, from (\ref{ii.V}) and (\ref{ii.VI}), we have the identity
\be
\lim_{q\to\epsilon}\left({(\[n\]_q!)^r\over\[rn+p\]_q!}
\right)^{1\over2}=
{q^{{-nrp\over2}}\over\sqrt{r!\[p\]_q!}}
\prod_{\alpha=0}^{r-1}q^{{n^2\alpha\over2}}\q.            \label{iii.XXII}
\ee
Note that we have used the sign ambiguity of the square root to choose the sign on the exponent in the  
$q^{-{nrp\over2}}$ term. This is of no physical significance, it just makes the equations tidier.
Using this and definitions (\ref{iii.XI}) and 
(\ref{iii.XVI}) we find that in the 
limit as $q\to\epsilon$, (\ref{iii.XXI}) becomes
\be\eqalign{
\lim_{q\to\epsilon}\vert rn+p\rangle
&=\lim_{q\to\epsilon}{(\ap)^p q^{-\frac{nrp}{2}}\over\sqrt{\[p\]_q!}
}\frac{1}{\sqrt{r!}}\left({(\ap)^n q^{{nN\over2}}\over \sqrt{\[n\]_q!}}\right)^r\vert0\rangle\q\cr
&={(\ap q^{-\frac{nN_b}{2}})^p\over\sqrt{\[p\]_q!}}
{(\bp)^r\over\sqrt{r!}}\vert0\rangle\q\cr
&={(\Ap)^p\over\sqrt{\[p\]_q!}}{(\bp)^r\over\sqrt{r!}}\vert0\rangle\q.        \label{iii.XXIII} 
}\ee 
Clearly this result means that we can write
\be
\lim_{q\to\epsilon}\vert rn+p\rangle=\vert r\rangle_{bosonic}
\otimes\vert p\rangle_{anyonic}\q.                                         \label{iii.XXIVa}
\ee 

This is the Fock space analogue of the algebraic decomposition which led to (\ref{iii.XVIII}).
Thus we see that for each generic $q$ state, there is a corresponding
state in the $q\to\epsilon$ limit. The difference is that when $q\to\epsilon$
these states are no longer part of a single $q$-oscillator irreducible representation, but are
instead in the product of a bosonic irreducible representation and an anyonic irreducible representation. Let us also
note that from (\ref{ii.IV}) and (\ref{iii.XXIII}), we know that for 
$q=\exp(\pm\frac{i\pi}{n})$ the states 
$\lim_{q\to\epsilon}\vert rn+p\rangle$ have positive definite norm 
(see \cite{DMAP2} for a discussion of this point).
In this case we have $(\bp)^\dagger=\bm$ and $(\Ap)^\dagger=\Am$, so that we can use the notation
\be
b^\dagger=\bp\q,\q b=\bm \q,\q A^\dagger=\ap\q,\q A=\am\q,                   \label{extra1}
\ee
since $\{b^\dagger,b,A^\dagger,A\}$ all have the implied hermiticity properties. 
It is particularly interesting to look at this result
when $q\to i$ (${\tilde n}$=4, $n$=2), since in this case the $q$-oscillator
decomposes into the 
physically observed bosonic and fermionic oscillators. To
see this we note that when $q=i$ we have 

\be 
AA^\dagger-i^{\pm1}A^\dagger A=i^{\mp\NA}\q,                                       \label{iii.XXV}
\ee
which acting on the Fock space, for which $p=0,1$
reduces to the familiar fermionic algebra,

\be
AA^\dagger+A^\dagger A=1\q.                                                         \label{iii.XXVI}
\ee

\section{Deformed angular momentum at $q$ a root of unity}
In this section we examine the structure of $\U_q(sl(2))$, and its subalgebra $\U_q(su(2))$, 
the enveloping algebra of the
$q$-deformed angular momentum algebra, in the \ql limit, with a view to finding structures 
analogous to those
found in association with fractional supersymmetry in
\cite{DMAP1,DMAP2,DMAP3,DMAP4,RSD}. For generic
$q$ the
algebraic part of $\U_q(sl(2))$ is

\be
[\sp,\sm]=\[2\sz\]\q,\hskip20pt q^{\sz}\spm q^{-\sz}=q^{\pm1}\spm\q,                  \label{iv.I}
\ee     
and it has coproduct, counit and antipode,

\be\eqalign{
\Delta\spm&=\spm\otimes q^{\sz}+q^{-\sz}\otimes\spm\q,\cr
\Delta q^{\pm\sz}&=q^{\pm\sz}\otimes q^{\pm\sz}\q,\cr
\varepsilon(\spm)&=0\q,\hskip15pt
\varepsilon(q^{\sz})=1\q,\cr
S(\spm)&=-q^{\pm1}\spm\q,\hskip15pt
S(q^{\sz})=q^{-\sz}\q.                                                            \label{iv.II}   
}\ee
$\U_q(sl(2))$ can be restricted to $\U_q(su(2))$ via the introduction of a \*-structure.
At algebraic level a \*-structure is an anti-involution satisfying
\be\eqalign{
*(\beta a)&=\beta^**(a)\q,\cr 
*^2(a)&=a\q,\cr
*(ab)&=*(b)*(a)\q.                                                        \label{iv.XV} 
}\ee
Here $\beta$ is a complex number with conjugate $\beta^*$ , and $a$ is an arbitrary element
of the algebra.
In the case of $\U_q(sl(2))$ we impose the following \*-structure, to make it into
$\U_q(su(2))$,
\be
*\spm=J_{\mp}\q,\q *q^{\pm\sz}=q^{\mp\sz}\q,                                 \label{iv.XVIb}     
\ee
which can be extended to the whole algebra using (\ref{iv.XV}).
 In matrix realizations of $\U_q(su(2))$ we can identify $*$ as hermitian
conjugation.
This \*-structure is compatible with the Hopf structure in the following sense
\be\eqalign{
\tau\circ\Delta\circ*&=*\otimes*\circ\Delta\q,\cr
\varepsilon\circ*&=*\circ\varepsilon\q,\cr
*\circ S\circ*&=S\q,                                                    \label{iv.XVIa}   
}\ee
where $\tau$ is the transposition map $\tau (X\otimes Y)=Y\otimes X$.

To avoid confusion we stress that the form of $\U_q(su(2))$ given here, and encountered in \cite{MajII} for example, is 
distinct from the more frequently encountered form of $\U_q(su(2))$, which has $q$ real,
a different $*$-structure, and compatibility relations which are different from those given in \eq{iv.XVIa}.

$\U_q(su(2))$ can be realized in terms of $q$-oscillators
by means of the $q$-deformed analogue of the 
Schwinger realization \cite{Mac1}. This realization
involves two copies $a_{1\pm}$ and $a_{2\pm}$ of the algebra
(\ref{iii.I}) which are mutually commutative, \ie\ $[a_{1\pm},a_{2(\pm)}]=0$,
and is explicitly defined by
\be
\sp=a_{1+}a_{2-}\q,\hskip10pt\sm=a_{1-}a_{2+}\q,\hskip10pt
 q^{\sz}=q^{N_1-N_2\over2}\q.                                       \label{iv.III}  
\ee
It is associated with those finite dimensional 
irreducible representations $\vert jm\rangle$ of $\U_q(sl(2))$, which can be defined
on the product of the Fock spaces of $a_{1\pm}$ and $a_{2\pm}$
as follows \cite{Mac1}
\be
\vert jm\rangle={(a_{1+})^{j+m}(a_{2+})^{j-m}\over
(\[j+m\]!\[j-m\]!)^{1\over2}}\vert0,0\rangle\q.                                 \label{iv.IIIa}
\ee
The action of the Schwinger realization (\ref{iv.III})
on these representations can be straightforwardly worked out
to be
\be
\eqalign{
\sm\vert jm\rangle&=(\[j+m\]\[j-m+1\])^{1\over2}\vert j,m-1\rangle\q,\cr
\sp\vert jm\rangle&=(\[j-m\]\[j+m+1\])^{1\over2}\vert j,m+1\rangle\q,\cr
q^{\pm\sz}\vert jm\rangle&=q^{\pm m}\vert jm\rangle\q.                   \label{iv.IIIb}
}\ee    
When \ql we can use (\ref{iii.XI}) to define the corresponding 
bosonic oscillators
\be
\eqalign{
\bop&=\qlm{(\aop)^n\over \sqrt{\[n\]!}}q^{{nN_1\over2}}\q,\q\hskip27pt
\bom=\qlm q^{{nN_1\over2}}{(\aom)^n\over \sqrt{\[n\]!}}\q,\cr
\btp&=\qlm{(\atp)^n\over \sqrt{\[n\]!}}q^{-{nN_2\over2}}\q,\q\hskip10pt
\btm=\qlm q^{-{nN_2\over2}}{(\atm)^n\over \sqrt{\[n\]!}}\q,                           \label{iv.IV}  
}\ee
and their number operators $N_{k b}=b_{k+} b_{k-}$ 
for $k=1,2$. Note that we have made a  change of sign $q^{{nN_2\over2}}\to q^{-{nN_2\over2}}$
in the second row, taking advantage of the freedom remarked upon previously. 
Using these we can construct 
a Schwinger realization of the undeformed $sl(2)$ algebra.
\be\eqalign{
\lp&=\bop\btm=\qlm{1\over\[n\]!}(\aop)^n(\atm)^n q^{{n(N_1-N_2+n)\over2}}\q\cr
&=\qlm{1\over\[n\]!}(\sp)^n q^{n\sz}q^{n^2\over2}\q,\cr
\lm&=\btp\bom=\qlm{1\over\[n\]!}q^{{n(N_1-N_2+n)\over2}}(\atp)^n(\aom)^n\q\cr               
&=\qlm{1\over\[n\]!}q^{n\sz}q^{n^2\over2}(\sm)^n\q,\cr  
\lz&={N_{b1}-N_{b2}\over2}={1\over2}[\lp,\lm]\q,                                          \label{iv.V}
}\ee 
so that
\be [\lp,\lm]=2\lz\q,\hskip20pt [\lz,\lpm]=\pm\lpm\q,                          \label{iv.VII} 
\ee                						
and also
\be
[\lz,\spm]=0\q,\q [q^{\sz},\lpm]_{q^n}\q.                                 \label{iv.VIII} 
\ee
As was the case with the $q$-oscillators, there is 
a more convenient basis for the algebra with generators $\{L_\pm,L_z,J_\pm,J_z\}$. We define
\be
q^{S_z}=q^{\sz-n\lz}\q,\quad S_{+}=q^{n\lz}\sp\q,\quad
 S_{-}=\sm q^{n\lz}\q,                                                 \label{iv.VIIII}
\ee  
so that 
\be
[S_+,S_-]=\[2S_z\]\q,\quad  [S_z,S_{\pm}]=\pm S_{\pm}\q,                       \label{iv.IX}    
\ee
and
\be
[\lz,S_{\pm}]=0\q,\q [q^{S_z},\lpm]=0\q.                                      \label{iv.X} 
\ee
In this basis the enveloping algebras $\U(sl(2))$
and $\U_q(sl(2))$ spanned by $\{\lpm,\lz\}$ and $\{S_\pm,q^{S_z}\}$
respectively are mutually commutative. In other words, 
we can write
\be
\qlm \U_q(sl(2))=\U_{\epsilon}(sl(2))\otimes \U(sl(2))\q,                        \label{iv.XI}
\ee
where by $\U_{\epsilon}(sl(2))$, we mean the enveloping algebra
obtained by simply setting $q=\epsilon$, rather than by taking the limit
as above.  

It is important to note that we have only established the
above decomposition for a particular realization.\par
In general the irreducible representations of an algebra are characterized
by the eigenvalues of its central elements. When $q^{\tilde n}=1$
, it follows directly from (\ref{iv.I}) that $\U_q(sl(2))$
has an expanded centre, now including $(\spm)^{{\tilde n}}$ and
$q^{4n\sz}$ as well as the usual Casimir operator. 
The structure of the representations is consequently
richer for such values of $q$. 
We now summarize the finite dimensional
irreducible representations of the $\U_q(sl(2))$ algebra for $q^{\tilde n}=1$.
For the derivation of this classification 
see \cite{Arnaudon1}. There are two basic cases.
Type A irreducible representations are labelled by two parameters, a half 
integral spin $j$ and a discrete parameter $\omega=\pm1,\pm i$.
They have dimension $2j+1$ where $0<2j+1\leq n$. If we use a basis
$\{\vert -j\rangle,\vert -j+1\rangle,...\vert j\rangle\}$, then
the action of the algebra on these irreducible representations is given
by
\be\eqalign{
\sm\vert m\rangle&=\omega(\[j+m\]\[j-m+1\])^{1\over2}\vert m-1\rangle\q,\cr
\sp\vert m\rangle&=\omega(\[j-m\]\[j+m+1\])^{1\over2}\vert m+1\rangle\q,\cr
\sm\vert -j\rangle&=0\q,\cr
\sp\vert j\rangle&=0\q,\cr
q^{\sz}\vert m\rangle&=\omega q^m\vert m\rangle\q,\cr
q^{-\sz}\vert m\rangle&=\omega^* q^{-m}\vert m\rangle\q.                           \label{iv.XII}
}\ee
Type B representations have dimension $n$, and are 
characterized by three complex parameters $j,x,y$.
If we use a basis 
$\{\vert -j\rangle,\vert -j+1\rangle,...\vert -j+n-1\rangle\}$,
then the action of the algebra on these irreducible representations is given by
\be\eqalign{
\sm\vert m\rangle&=(\[j+m\]\[j-m+1\]+xy)^{1\over2}\vert m-1\rangle\q,\cr
\sp\vert m\rangle&=(\[j-m\]\[j+m+1\]+xy)^{1\over2}\vert m+1\rangle\q,\cr
\sm\vert -j\rangle&=y\vert -j+n-1\rangle\q,\cr
\sp\vert -j+n-1\rangle&=x\vert -j\rangle\q,\cr
q^{\sz}\vert m\rangle&= q^m\vert m\rangle\q.                       \label{iv.XIII}
}\ee
When $x,y\neq0$ the irreducible representations \eq{iv.XIII} are said to be cyclic,
and when $x=0$, $y\neq0$ or $y=0$, $x\neq0$ they are
said to be semicyclic.
 Two independent restrictions which we can impose upon
$\U_q(sl(2))$, both of them compatible with the coproduct are
\be
\sp^n=0\q,\hskip20pt \sm^n=0\q.                                          \label{iv.XIV} 
\ee
When these restrictions are imposed the algebra plainly loses 
the cyclic and semicyclic irreducible representations from (\ref{iv.XIII}). When the \*-structure
(\ref{iv.XVIb}) is also imposed the only remaining irreducible representations are 
those type A irreducible representations with $\omega=\pm1$. 
For these irreducible representations we have
\be
\spm^\dagger=J_{\mp}\q,\q (q^{\pm\sz})^\dagger=q^{\mp\sz}\q,                     \label{iv.XVII}
\ee  
which reflects the \*-structure (\ref{iv.XVIb}). We will refer 
to this restricted \*-Hopf algebra as $\U_q(su(2),r)$. Note also that on any irreducible representation
$q^{\sz}=\omega q^m$ where $m$ takes on half integer values in the range $-j$ to $+j$, $2j+1\leq n$. 
Consequently $q^{\sz}$ satisfies the characteristic polynomial
\be
\prod_{m=0}^{n-1}(q^{4n\sz}-q^{2m})=0\q,                                  \label{iv.XIVa}
\ee
which is equivalent to both
\be
\prod_{m=0}^{n-1}\[2\sz-m\]_q=0\q,                                 \label{iv.XIVb}
\ee     
and $q^{4n\sz}=1$. 
Apart from the $\omega$ factor, the irreducible representations (\ref{iv.XII}) are identical to those associated with 
the Schwinger realization (\ref{iv.IIIb}). 
For $\U_q(sl(2))$ we cannot use the $*$-structure to restrict the algebra, and instead
introduce $q^{4n\sz}=1$ directly. This condition also restricts us to the type A irreducible representations, 
only now $\omega$ can take on any of the values $\omega=\pm1,\pm i$. For more on restricted forms of $\U_q(sl(2))$ and other $\U_q(g)$ enveloping algebras see \cite{MajII,CP} and the references therein. 
The results we obtained using the 
Schwinger realization suggest that there may be some extra structure in the $q\to\epsilon$ limit, 
the part associated with $\lpm$.
Motivated by ideas introduced in \cite{LUSZTIG1,LUSZTIG2}, and the results of \cite{DMAP1,DMAP2,DMAP3,DMAP4,RSD} 
we now look for a FSUSY-like extension of $\U_q(sl(2),r)$ which includes this extra structure.
In what follows everything that is said about $\U_q(sl(2),r)$, 
holds equally for $\U_q(su(2),r)$, the reality properties of which will also be worked out .
At generic $q$, the Hopf algebraic structure of the quantities
\be
{\spm^n\over\[n\]!}\q,\q J_z\q,                                                 \label{iv.XVIII}
\ee
is well defined. We now extend $\U_q(sl(2),r)$ by the elements $\spm^{(n)},J_z$
to make $\U_q(sl(2),f)$ ($f$ for fractionally supersymmetric). $\spm^{(n)},J_z$ are to be endowed with the Hopf algebraic structure
associated with the \ql limit of ${\spm^n\over\[n\]!},J_z$, so that they add to $\U_q(sl(2),r)$
some more of the structure associated with the generic $q$ case. We do not make the explicit
identification $\spm^{(n)}=\qlm{\spm^n\over\[n\]!}\q,$ since the latter quantity is not well
defined algebraically (although it is for all representations). 
From \cite{GRS}, we have
the identities
\be
[\sp^m,\sm^s]=\sum_{p=1}^{min(m,s)}
\left[\!\!\left[\matrix{m\cr p}\right]\!\!\right]
\left[\!\!\left[\matrix{s\cr p}\right]\!\!\right]\[p\]!
\sm^{s-p}\left(\prod_{k=p-m-s+1}^{2p-m-s}\[2\sz+k\]\right)\sp^{m-p}\q.         \label{iv.XIX}
\ee
We can use them to determine the algebraic properties of $\spm^{(n)}$, which
we obtain by taking the \ql limits of the algebraic properties of ${\spm^n\over\[n\]!}$,
\be
\eqalign{
[J_+^{(n)},J_-]&=q^{-n}{\[2J_z+1\]J_+^{n-1}\over\[n-1\]!}\q,\cr
[J_+,J_-^{(n)}]&=q^{-n}{J_-^{n-1}\[2J_z+1\]\over\[n-1\]!}\q.                    \label{iv.XX} 
}\ee
In addition there are the trivial results $[\spm^{(n)},J_{\pm}]=0$.   
We also have
\be
[J_+^{(n)},J_-^{(n)}]=\sum_{p=1}^{n-1}{1\over(\[n-p\]!)^2\[p\]!}
\sm^{n-p}\left(\prod_{k=p+1}^{2p}\[2\sz+k\]\right)\sp^{n-p}
+\qlm{1\over\[n\]!}\prod_{k=1-n}^{0}\[2\sz+k\]\q.                                      \label{iv.XXIa}
\ee    
As a consequence of (\ref{iv.XIVb}), the last term in this equation is 
algebraically well defined. Its explicit value, which is worked out in
the appendix, is
\be
\qlm{1\over\[n\]!}\prod_{k=1-n}^0\[2J_z+k\]=
{q^{2n\sz-n^2}\left({2\sz-\frac{3n}{2}-\frac{1}{2}\over n}\right)}
+{q^{2n\sz}q^{-{n(n+1)\over2}}\over(1-q^{-2})^n}\sum_{k=1}^{n-1}{(-q^{(1-4\sz+n)})^k
\over\[n-k\]!\[k\]!}\q,                                                          \label{iv.XXIb} 
\ee   

and we will make use of this below.
The commutation relations of $J_z$ are easily found to be
\be
[\sz,\spm]=\pm\spm\q,\quad [\sz,\spm^{(n)}]=\pm n\spm^{(n)}\q.                       \label{iv.XXI}  
\ee 
The Hopf structure given by (\ref{iv.II}), (\ref{iv.XIV}) and $q^{4nJ_z}=1$ leads 
in a similar fashion to the following Hopf structure for $J_{\pm}^{(n)},\sz$,  
\be
\eqalign{
\Delta J_{\pm}^{(n)}&=J_{\pm}^{(n)}\otimes q^{nJ_z}+q^{-nJ_z}\otimes J_{\pm}^{(n)}
+\sum_{k=1}^{n-1}{q^{(k-n)J_z}J_\pm^k\otimes 
q^{kJ_z} J_\pm^{n-k}\over\[k\]!\[n-k\]!}\q,\cr
\Delta \sz&=1\otimes\sz+\sz\otimes1\q,\cr                                                                       
S(J_{\pm}^{(n)})&=-q^{-n^2}J_{\pm}^{(n)}\q,\quad\quad S(J_z)=-J_z\q,\cr
\varepsilon(J_{\pm}^{(n)})&=0\q,\quad\hskip46pt\varepsilon(\sz)=0\q,                   \label{iv.XXII}
}\ee
and for $\U_q(su(2),f)$ the $*$-structure 
\be
*(J_{\pm}^{(n)})=J_{\mp}^{(n)}\q,\hskip21pt *(\sz)=\sz\q.                   \label{iv.XXIII}
\ee
Equations (\ref{iv.XX})-(\ref{iv.XXII}) show that as expected
 the additional elements $J_{\pm}^{(n)},\sz$ can be added to the Hopf 
algebra without contradiction, so that $\U_q(sl(2),f)$ is well defined
as a Hopf algebra. When \eq{iv.XXIII} is also imposed, the corresponding result for $\U_q(su(2),f)$ is obtained.
 $\U_q(sl(2),r)$ is a sub-Hopf algebra of $\U_q(sl(2),f)$,
for which we already know all of the irreducible representations (\ref{iv.XII}). Let us denote a state in an highest weight representation of
 $\U_q(sl(2),f)$ by $\vert m,?\rangle$,
in which the $m$ is associated with the $\U_q(sl(2),r)$ subalgebra, and the ? denotes any, as yet unknown, extra labeling, which is needed to fully describe the 
highest weight representations of the extended algebra. Using (\ref{iv.XX}) and (\ref{iv.XI}) we find
\be\eqalign{
[\sp^{(n)},\sm]\vert m,?\rangle
&=q^n{\[2J_z+1\]J_+^{n-1}\over\[n-1\]_q!}\vert m,?\rangle\q\cr
&=0\q,                                                                   \label{iv.XXIV} 
}\ee
and similarly
\be
[J_+,J_-^{(n)}]\vert m,?\rangle=0\q,                                         \label{iv.XXV}
\ee
as well as the trivial results,
\be
[J_{\pm},J_{\pm}^{(n)}]\vert m,?\rangle=0\q.                             \label{iv.XXVI}      
\ee
Note that these results hold regardless of the full form of $\vert m,?\rangle$,
since they depend only on the $m$ part. Using (\ref{iv.XX}) and (\ref{iv.XXIV})-(\ref{iv.XXVI})
we can also evaluate the following without knowing the full form of $\vert m,?\rangle$.
\be\eqalign{
[[\sp^{(n)},\sm^{(n)}],\sp^{(n)}]\vert m,?\rangle
&=\left[\sum_{p=1}^{n-1}{1\over(\[n-p\]!)^2\[p\]!}
\sm^{n-p}\left(\prod_{k=p+1}^{2p}\[2\sz+k\]\right)\sp^{n-p},\sp^{(n)}\right]\vert m,?\rangle\cr
&+\left[q^{2n\sz+n^2}\left({2\sz-\frac{3n}{2}-\frac{1}{2}\over n}\right),\sp^{(n)}\right]\vert m,?\rangle\cr
&+\left[{q^{2n\sz}q^{-{n(n+1)\over2}}\over(1-q^{-2})^n}\sum_{k=1}^{n-1}{(-q^{(1-4\sz+n)})^k
\over\[n-k\]!\[k\]!},\sp^{(n)}\right]\vert m,?\rangle\q.              \label{iv.XXVII}   
}\ee
From (\ref{iv.XXIV})-(\ref{iv.XXVI}) the first and third terms term are zero.
The second term then yields, 
\be
[[\sp^{(n)},\sm^{(n)}],\sp^{(n)}]\vert m,?\rangle
=2\sp^{(n)}q^{2n\sz+n^2}\vert m,?\rangle\q.                                 \label{iv.XXVIII}   
\ee
Similarly we find that
\be
[[\sp^{(n)},\sm^{(n)}],\sm^{(n)}]\vert m,?\rangle
=-2\sm^{(n)}q^{2n\sz+n^2}\vert m,?\rangle\q.                                        \label{iv.XXIX} 
\ee  
These results have their most natural form in the $L,S$ basis
introduced earlier. We can change to this using
\be
\lp=\sp^{(n)}q^{n\sz}q^{n^2\over2}\q,\hskip20pt 
\lm=q^{n\sz}q^{n^2\over2}\sm^{(n)}\q,\hskip20pt 
\lz={[\lp,\lm]\over2}\q,                                                       \label{iv.XXX}
\ee
and
\be
S_+=q^{n\lz}\sp\q,\hskip20pt 
S_-=\sm q^{n\lz}\q,\hskip20pt 
q^{S_z}=q^{J_z-nL_z}\q,                                                     \label{iv.XXXI}    
\ee
which follow from (\ref{iv.V}) and (\ref{iv.VIIII}).
Note that from (\ref{iv.XVIb}) and (\ref{iv.XXIII}), the \*-structure of the elements in $\U_q(su(2),f)$, the \*-structure of the elements in this modified  basis is
\be
\eqalign{
*(\lp)&=q^{-2nS_z-2n^2L_z-n^2}L_-\q, \hskip20pt *(S_+)=S_-q^{-2n\lz}\q,\cr
*(\lm)&=L_+q^{-2nS_z-2n^2L_z-n^2}\q, \hskip20pt
*(S_-)=q^{-2n\lz}S_+\q,\cr 
*(L_z)&=q^{-4n(S_z+nL_z)}L_z\q,\hskip46pt 
*(q^{S_z})=q^{-S_z}q^{-n(L_z-q^{-4n(S_z+nL_z)}L_z)}\q.     \label{iv.XXXII}
}\ee
Although the algebra described by (\ref{iv.XX})-(\ref{iv.XXI}) does not have the direct product form seen in
(\ref{iv.XI}), it follows none the less from 
(\ref{iv.XXIV})-(\ref{iv.XXIX}) that for all highest weight  representations,
\be
\eqalign{
[S_+,S_-]&=\[2S_z\]\q,\hskip20pt  [S_z,S_{\pm}]=\pm S_{\pm}\q,\cr
[\lp,\lm]&=2\lz\q,\hskip25pt [\lz,\lpm]=\pm\lpm\q,\cr
[S_{\pm},\lpm]&=0\q,\hskip38pt [S_z,\lpm]=0\q,\hskip20pt[\lz,S_{\pm}]=0\q.    \label{iv.XXXIII}
}\ee 
In consequence, the highest weight representations of $\U_q(sl(2),f)$ are a direct product of the form
\be
\U_q(sl(2),r)\otimes \U(sl(2))\q,            \label{dirpro} 
\ee
in which $\U(sl(2))$ denotes the enveloping algebra of undeformed
$sl(2)$. It remains to establish the equivalent results for $\U_q(su(2),f)$ with the reality 
properties implied by \eq{iv.XXXII}.There are three basic cases:\par
i) ${\tilde n}$ odd. In this case the $*$-structure \eq{iv.XXXII} implies the following hermiticity
properties for the algebraic elements on Fock space representations.
\be
S_\pm^\dagger=S_\mp\q,\q (q^{S_z})^\dagger=q^{-S_z}\q,
\q L_\pm^\dagger=L_\mp\q,\q L_z^\dagger=L_z\q,                          \label{KK1}         
\ee
so that the analogue of \eq{dirpro} is 
\be
\U_q(su(2),r)\otimes \U(su(2))\q.            \label{KK2} 
\ee
If we take $q=\exp(\frac{2a\pi i}{n})$ with $a$ an integer, 
then since $\[m\]$ goes negative for $m>\frac{n}{2a}$ only the 
$\U_q(su(2),r)$ irreducible representations with $2j_s+1<\frac{n}{2a}$, \ie\ those with $j<\frac{n-2a}{4a}$ are unitary. The
simplest non-trivial example has $a=1$, ${\tilde n}=n=5$, for which this unitarity condition gives
$j_s<\frac{3}{4}$, \ie\ $j_s=0,\frac{1}{2}$. The $j_s=0$ irreducible representation is trivial, and the action of the
algebraic elements on the $j_s=\frac{1}{2}$ irreducible representation is given by
\be\eqalign{
\sp\left\vert-{1\over2},{1\over2}\right\rangle&=\left\vert{1\over2},{1\over2}\right\rangle\q,\hskip20pt
\sm\left\vert-{1\over2},{1\over2}\right\rangle=0\q,\cr
\sm\left\vert{1\over2},{1\over2}\right\rangle&=\left\vert-{1\over2},{1\over2}\right\rangle\q,\hskip20pt
\sp\left\vert{1\over2},{1\over2}\right\rangle=0\q,                                 \label{KK3}
}\ee 
which is identical to the undeformed  $s={1\over2}$ irreducible representation. Thus \eq{KK2} can be interpreted as a
single wave
function consisting of the direct product of an intrinsic spin degree of freedom with an orbital
angular momentum part. For higher odd ${\tilde n}$ we can extend this interpretation so that in general \eq{KK2}
can be viewed as
\be
{\rm Intrinsic\hskip5pt anyonic\hskip5pt spin}
\otimes{\rm Orbital\hskip5pt angular\hskip5pt momentum}.             \label{KK4} 
\ee
Note that to make \eq{KK4} strictly accurate we have to exclude non integer values of $j_l$. This can be done in the 
usual way by imposing $\psi(\theta+2\pi)=\psi(\theta)$ on the wave functions.\par
ii) ${\tilde n}$ even, $n$ even. In this case the hermiticity properties implied by \eq{iv.XXXII} are
\be
\eqalign{
S_\pm^\dagger&=(-1)^{2L_z}S_\mp\q,\q  (q^{S_z})^\dagger=q^{S_z}\q,\cr 
L_\pm^\dagger&=(-1)^{2S_z}L_\mp\q,\q  L_z^\dagger=L_z\q,                 \label{KK5} 
}\ee
so that in this case the analogue of \eq{dirpro} is
\be
\eqalign{
\U_q(su(2))&\otimes\U(su(2))\hskip30.5pt {\rm for\hskip5pt 2{\it j_s}\hskip5pt even,\q 2{\it j_l}\hskip5pt even},\cr
\U_q(su(2))&\otimes\U(su(1,1))\hskip20pt {\rm for\hskip5pt 2{\it j_s}\hskip5pt odd,\q 2{\it j_l}\hskip5pt even},\cr
\U_q(su(1,1))&\otimes\U(su(2))\hskip30.5pt {\rm for\hskip5pt 2{\it j_s}\hskip5pt even,\q 2{\it j_l}\hskip5pt odd},\cr
\U_q(su(1,1))&\otimes\U(su(1,1))\hskip20pt {\rm for\hskip5pt 2{\it j_s}\hskip5pt odd,\q 2{\it j_l}\hskip5pt odd}.     
\label{KK5.5}
}\ee
It follows from \eq{iv.XIII} that for $2j_l$ odd 
\be
S_+S_-\vert m_s\rangle=-\[j_s-m_s\]\[j_s+m_s+1\]\vert m_s\rangle\q,         \label{KK6}
\ee
so that there are always negative norm states in these irreducible representations, and consequently they are not physical (\ie\ not unitary).
On the other hand when $2j_l$ is even, and with $q=\exp(\pm {\pi i\over n})$, all of the irreducible representations
are unitary (for other values of $q$ there are restrictions on the value of $j_s$ as in case (i)). Thus the unitary highest weight representations of $\U_q(su(2),f)$ all have integer $j_l$. In this case the interpretation of the $L$ part as orbital angular momentum follows without the need for any further restrictions. Note that there are also highest weight representations in which the $L$ part is a $\U(su(1,1))$ algebra with integer $j_l$. These correspond to orbital angular momentum in a space of $1+2$ dimensions.

\par
iii) ${\tilde n}$ even, $n$ odd. In this case the hermiticity properties implied by \eq{iv.XXXII} are
\be
\eqalign{
S_\pm^\dagger&=(-1)^{2L_z}S_\mp\q,\q  (q^{S_z})^\dagger=q^{S_z}\q,\cr 
L_\pm^\dagger&=(-1)^{2S_z+2L_z+1}L_\mp\q,\q  L_z^\dagger=L_z\q,                 \label{KK7} 
}\ee
so that in this case the analogue of \eq{dirpro} is
\be
\eqalign{
\U_q(su(2))&\otimes\U(su(2))\hskip20pt \hskip11pt{\rm 
for\hskip5pt 2{\it j_s}\hskip5pt odd,\q 2{\it j_l}\hskip5pt even},\cr
\U_q(su(2))&\otimes\U(su(1,1))\hskip20pt \hskip1pt{\rm 
for\hskip5pt 2{\it j_s}\hskip5pt even,\q 2{\it j_l}\hskip5pt even},\cr
\U_q(su(1,1))&\otimes\U(su(2))\hskip20pt \hskip11pt{\rm 
for\hskip5pt 2{\it j_s}\hskip5pt even,\q 2{\it j_l}\hskip5pt odd},\cr
\U_q(su(1,1))&\otimes\U(su(1,1))\hskip20pt \hskip2pt{\rm 
for\hskip5pt 2{\it j_s}\hskip5pt odd,\q 2{\it j_l}\hskip5pt odd},     
\label{KK8}
}\ee
As for case (ii), only the first two classes of highest weight 
representation in the list given above have members which are unitary.\par
In \cite{LUSZTIG1,LUSZTIG2} a more rigorous and elegant, though less direct,
approach to some of the material covered here is introduced in a
purely mathematical context. In particular, this method leads to a
generalization of the  results in this  section to all $\U_q(g)$
enveloping algebras. 
Let us conclude this section by noting that the direct 
product structure of the highest weight representations 
of $\U_q(sl(2))$ at $q=\epsilon$ was anticipated by the 
result \eq{iv.XI} for the deformed Schwinger realization. 
There are deformed bosonic realizations of all $\U_q(g)$ 
enveloping algebras \cite{Hayashi}, for which at 
$q=\epsilon$, similar decompositions are to be expected. 
Perhaps such realizations will serve as  useful tools for 
studying fractionally supersymmetric extensions of $\U_q(g)$.

\section{$GL_q(2)$ at $q$ a root of unity}

The quantum group $SL_q(2)$ is dual \cite{FRT} to $\U_q(sl(2))$, and so it is reasonable to 
expect that there are analogous extensions and decompositions for this Hopf algebra when 
$q$ is a root unity. 
To begin with we will work with the complete $GL_q(2)$ Hopf algebra. This quantum group
is generated by the elements $\{a,b,c,d\}$ of the matrices
\be
\omw=\left(\matrix{a &b\cr c &d}\right)\q,                     \label{v.II}
\ee
which have the nontrivial commutation relations given below
\be
\eqalign{
[a,b]_q&=0\q,\quad[a,c]_q=0\q,\quad [a,d]=\lambda bc\q,\cr
[b,c]&=0\q,\quad [b,d]_q=0\q,\quad[c,d]_q=0\q,                          \label{v.III}
}\ee
with $\lambda=q-q^{-1}$.
The coproduct is
\be
\eqalign{
\Delta(a)&=a\otimes a+b\otimes c\q,\quad\Delta(b)=a\otimes b+b\otimes d\q,\cr
\Delta(c)&=c\otimes a+d\otimes c\q,\quad\Delta(d)=c\otimes b+d\otimes d\q,    \label{v.IV}
}\ee
and the counit and antipode are given by
\be
\eqalign{
\varepsilon(a)&=\varepsilon(d)=1\q,\quad \varepsilon(b)=\varepsilon(c)=0\q,\cr
S(a)&=(\det_q\omw)^{-1}d\q,\quad S(b)=-q^{-1}(\det_q\omw)^{-1}b\q,\cr
S(c)&=-q(\det_q\omw)^{-1}c\q,\quad S(d)=(\det_q\omw)^{-1}a\q,                  \label{v.V}   
}\ee
where
\be
\det_q\omw:=ad-qbc\q,                                                 \label{v.VI}
\ee
is a central and grouplike element known as the `quantum determinant' of $\omw$. 
Here the word grouplike is used to indicate that $\Delta
g=g\otimes g$ for $g=\det_q\omw$. 
$GL_q(2)$ can be restricted to $SL_q(2)$ by setting $\det_q\omw=1$. Two Hopf algebras are said to
be dually paired if
\be\eqalign{
\langle xy,\alpha\rangle&=\langle x\otimes y,\Delta\alpha\rangle\q,\cr
\langle x,\alpha\beta\rangle&=\langle \Delta x,\alpha\otimes\beta\rangle\q,\cr
\langle 1,\alpha\rangle&=\varepsilon(\alpha)\q,\cr
\langle x,1\rangle&=\varepsilon(x)\q,\cr
\langle S(x),\alpha\rangle&=\langle x,S(\alpha)\rangle\q.            \label{v.VIa}
}\ee
This duality can be extended to the \*-structure in more than one way \cite{MajII,STAR}.
For our purposes the most convenient form is
\be
\langle x,\alpha\rangle=\langle *x,*\alpha\rangle^*\q.              \label{v.VIb}
\ee
This pairing is often degenerate.
For $SL_q(2)$ and $\U_q(sl(2))$ there is a pairing of type (\ref{v.VIa})
given by
\be
\eqalign{
\langle q^{\pm J_z},a\rangle&=\langle q^{\pm J_z},d\rangle=q^{\pm{1\over2}}\q,\cr
\langle q^{\pm J_z},b\rangle&=\langle q^{\pm J_z},c\rangle=1\q,\cr
\langle \sp,b\rangle&=\langle \sm,c\rangle=1\q,\cr
\langle \sp,a\rangle&=\langle \sm,d\rangle=0\q,                         \label{v.VIc} 
}\ee
and extended to products using (\ref{v.VIa}). For details see \cite{FRT,MajII,CP}.
At generic $q$, we have
\be
\eqalign{
\langle \sp^n,b^n\rangle&=[n]_{q^2}!=q^{n(n-1)}\[n\]!\q,\cr
\langle \sm^n,c^n\rangle&=[n]_{q^{-2}}!=q^{-n(n-1)}\[n\]!\q,                            \label{v.VIaa}
}\ee
which are easily found using induction. Note that $b^n,c^n$ from $SL_q(2)$ are not paired to any elements in $\U_q(sl(2))$ other than $J^n_+$ and $J^n_-$ respectively, so that when $q=\epsilon$, they have a null pairing with all of $U_q(sl(2))$.
There are two interesting ways of retaining the non-null pairing from the generic case.
Our work in section 4, with $\U_q(su(2),f)$ provides one. In this case the null pairing is a
straightforward consequence of $\spm^n=0$, so that by rearranging (\ref{v.VIaa}) and taking the $q\to\epsilon$ limit
we obtain the following
pairings for $\spm^{(n)}$,
\be\eqalign{
\langle \sp^{(n)},b^n\rangle&=q^{n(n-1)}\q,\cr 
\langle \sm^{(n)},c^n\rangle&=q^{-n(n-1)}\q,                                 \label{v.VIab}   
}\ee
and also find that they have null pairings with the rest of $SL_q(2)$. An immediate consequence of
this is that the form of $SL_q(2)$ dual to $\U_q(su(2,f))$ has $b^n,c^n\neq0$.  
A second way of retaining
the non-null generic $q$ pairings will be discussed later in this section.
From (\ref{v.VIc}) it follows immediately that
\be\eqalign{
\langle ad-qbc,1\rangle&=\langle a\otimes d-q b\otimes c,1\otimes 1\rangle\q\cr
&=\langle a,1\rangle\langle d,1\rangle-q\langle b,1\rangle\langle c,1\rangle\cr  
&=1\q,                                                                              \label{v.VId}
}\ee
which is compatible with $\det_q\omw=1$. 
If we impose the $*$-structure (\ref{iv.XVIb}) on 
$\U_q(sl(2))$ to make it into $\U_q(su(2))$ then from (\ref{v.VIb}) we obtain
the following hermitian \*-structure on $\omega$
\be
*\left(\matrix{a&b\cr c&d}\right)=\left(\matrix{a&c\cr b&d}\right)\q.            \label{v.VIe}       
\ee
It is easy to check that this is compatible with (\ref{iv.XV}) and (\ref{iv.XVIa}).
We will refer to $SL_q(2)$ with this \*-structure as $*SL_q(2)$ and to
$GL_q(2)$ with the same \*-structure as $*GL_q(2)$. Note that although 
$*SL_q(2)$ is dual to our $\U_q(su(2))$, it is not the quantum group usually
referred to as $SU_q(2)$, which involves real $q$ and a different \*-structure. 
Let us now define
\be
A=a^n\q,\quad B=b^n\q,\quad C=c^n\q,\quad D=d^n\q.                            \label{v.VII}
\ee     
From (\ref{v.III}) it follows directly that
\be
\eqalign{
[A,b]_{q^{n}}&=[A,c]_{q^{n}}=[D,b]_{q^{n}}=[D,c]_{q^{n}}=0\q,\cr
[B,a]_{q^{n}}&=[B,d]_{q^{n}}=[C,a]_{q^{n}}=[C,d]_{q^{n}}=0\q,\cr
[A,d]&=[D,a]=[B,c]=[C,b]=0\q,                                         \label{v.VIII}    
}\ee 
in regard to which we recall that $q^{n}=\pm1$. Also
\be
\eqalign{
[A,B]_{q^{n^2}}&=0\q,\quad[A,C]_{q^{n^2}}=0\q,\quad [A,D]=0\q,\cr
[B,C]&=0\q,\quad [B,D]_{q^{n^2}}=0\q,\quad[C,D]_{q^{n^2}}=0\q,          \label{v.IX}
}\ee
where likewise it should be noted that $q^{n^2}=\pm1$.
Let us now go on to consider the rest of the Hopf structure. From (\ref{v.IV}),
the coproduct of $A$ is given by
\be
\eqalign{
\Delta (A)&=\Delta (a^n)=(a\otimes a+b\otimes c)^n\q\cr
&=a^n\otimes a^n+b^n\otimes c^n+\sum_{m=1}^{n-1}
\left[\!\!\left[\matrix{n\cr m\cr}\right]\!\!\right]q^{m^2-nm}(a\otimes a)^{n-m}(b\otimes c)^m\q\cr
&=a^n\otimes a^n+b^n\otimes c^n=A\otimes A+B\otimes C\q.          \label{v.X}
}\ee 
The coproducts of $B,C$ and $D$ can be similarly derived, and we find that
\be
\eqalign{
\Delta (A)&=A\otimes A+B\otimes C\q,\quad \Delta(B)=A\otimes B+B\otimes D\q,\cr
\Delta (C)&=C\otimes A+D\otimes C\q,\quad \Delta(D)=C\otimes B+D\otimes D\q,   \label{v.XI}
}\ee
which has the same form as (\ref{v.IV}). The counit and antipode of 
$\{A,B,C,D\}$ are easily deduced to be
\be
\varepsilon(A)=\varepsilon(D)=1\q,\quad \varepsilon(B)=\varepsilon(C)=0\q,                 \label{v.XII}
\ee     
and
\be
\eqalign{
S(A)&=(\det_q\omw)^{-n}D\q,\hskip49pt S(B)=-q^{n^2}(\det_q\omw)^{-n}B\q,\cr
S(C)&=-q^{-n^2}(\det_q\omw)^{-n}C\q,\quad S(D)=(\det_q\omw)^{-n}A\q.          \label{v.XIII} 
}\ee
Writing
\be
\eqalign{
AD&=a^nd^n\q\cr
  &=a^{n-1}(\det_q\omw+qbc)d^{n-1}\q\cr
  &=\prod_{r=1}^n(\det_q\omw+(q^{-1}bc)q^{2r})^n\q,                        \label{v.XIV} 
}\ee
and using the identity \cite{DMAP2}
\be
\prod_{r=1}^n (\alpha+p^r\beta)=\alpha^n+p^{{n(n+1)\over2}}\beta^n\q,             \label{v.XV}
\ee
in which $p$ is a root of unity,
we have
\be
\eqalign{
AD&=(\det_q\omw)^n+q^{n(n-1)}q^{-n}b^nc^n\q\cr
  &=(\det_q\omw)^n+q^{n^2}BC\q.                                              \label{v.XVI}   
}\ee
Now, if we define
\be
W:=\left(\matrix{A &B\cr C &D}\right)\q,                               \label{v.XVII}
\ee
so that
\be
\det_{q^{n^2}}W=AD-q^{n^2}BC\q,                                                 \label{v.XVIII} 
\ee
we have
\be
(\det_q\omw)^n=\det_{q^{n^2}}W.                                                  \label{v.XIX}
\ee
Using this, we can rewrite (\ref{v.XIII}) as
\be
\eqalign{
S(A)&=(\det_{q^{n^2}}W)^{-1}D\q,\hskip49pt 
S(B)=-q^{n^2}(\det_{q^{n^2}}W)^{-1}B\q,\cr
S(C)&=-q^{-n^2}(\det_{q^{n^2}}W)^{-1}C\q,\quad 
S(D)=(\det_{q^{n^2}}W)^{-1}A\q.                                                \label{v.XX} 
}\ee   
By comparing (\ref{v.IX}),(\ref{v.XI}),(\ref{v.XII}), (\ref{v.XVIII}) 
and (\ref{v.XX}) with (\ref{v.III})-(\ref{v.VI}), we see that the elements
$\{A,B,C,D\}$ generate a $GL_{q^{n^2}}(2)$ sub-Hopf algebra of $GL_q(2)$. 
In the case of odd ${\tilde n}$, $q^{n^2}=1$, so that this sub-Hopf algebra 
is just undeformed $GL(2)$, and moreover since $q^n=1$, it is central.
We can specialize to $SL_q(2)$ by fixing $\det_q\omw=1$. From (\ref{v.XIX}) this
implies that $\det_{q^{n^2}}W=1$, so that the subalgebra generated by $\{A,B,C,D\}$ is itself
restricted to $SL_{q^{n^2}}(2)$. Similarly, when we impose the \*-structure (\ref{v.VIe}) we find
from (\ref{v.VII}) that a \*-structure of the same form is induced on $W$, \ie\
\be
*\left(\matrix{A&B\cr C&D}\right)=\left(\matrix{A&C\cr B&D}\right)\q,            \label{v.XXa}       
\ee
so that $*GL_q(2)$ and $*SL_q(2)$ have, respectively, sub-Hopf algebras $*GL_{q^{n^2}}(2)$
and $*SL_{q^{n^2}}(2)$.
The following observations are intended to assist in the development of a physical
interpretation of these results. \vskip0pt\noindent
i) The \*-structure preserves the determinant, \ie\
\be
\eqalign{
*\det_q\omw&=\det_q\omw\q,\cr *\det_{q^{n^2}}W&=\det_{q^{n^2}}W\q.      \label{v.XXaa}  
}\ee
\vskip0pt\noindent
ii) The theory of covariant transformations of $GL_q(2)$, has been discussed by several authors,
\eg\ \cite{mink1,mink2,mink3,mink4}.
As one would expect, these preserve the determinant $\det_q\omw$, and thus from (\ref{v.XIX})
$\det_{q^{n^2}}W$ as well.
\vskip0pt\noindent
iii) If we write
\be
\omw=\left(\matrix{ p_0-p_3&p_1-ip_2\cr p_1+ip_2&p_0+p_3\cr}\right)
=\left(\matrix{a&b\cr c&d}\right)\q,                                            \label{v.XXb}
\ee
and
\be
W=\left(\matrix{P_0-P_3&P_1-iP_2\cr P_1-iP_2&P_0+P_3\cr}\right)
=\left(\matrix{A&B\cr C&D}\right)\q,                                        \label{v.XXc}
\ee
then the reality of $\{p_\mu\}$ and $\{P_\mu\}$ follow from the
\*-structure (\ref{v.VIe}) and (\ref{v.XXa}). Also, from (\ref{v.III}).
\be
\eqalign{
[p_0,p_3]&={1\over4}[a+d,d-a]={1\over2}[a,d]\q,\cr
&={1\over2}\lambda bc={1\over2}\lambda (p_2^2+p_1^2)\q,                             \label{v.XXd}
}\ee
so that 
\be
\eqalign{
\det_q\omw&=p_0^2-p_3^2+[p_0,p_3]-q(p_2^2+p_1^2)\q\cr
&=p_0^2-{q+q^{-1}\over2}(p_2^2+p_1^2)-p_3^2\q,\cr                                     \label{v.XXe}
}\ee
and similarly
\be
\det_{q^{n^2}}W=P_0^2-{q^{n^2}}(P_2^2+P_1^2)-P_3^2\q.                             \label{v.XXf}
\ee     
iv) When $q=1$, we have $\omw=W$, and $\det W$ is the Laplacian on
undeformed Minkowski space, which appears in the Klein-Gordon equation, \ie\ ($\det_{q}w^2-M^2)\vert\psi\rangle=0$
.\par
Based on these observations we interpret $*GL_q(2)$ as the algebra of quantized momenta (\ie\
derivatives up to a factor of $i$) on $q$-deformed Minkowski-space, with $\det_q\omw$ as the
$q$-deformed Laplacian.  A central feature of this deformed momentum-space
is that for $q^{n^2}=1$ it contains undeformed Minkowski momentum-space, \ie\ $*GL(2)$ with coordinate functions 
$\{P_i\}$ as a sub-Hopf algebra. To make this interpretation more explicit we introduce the notation
\be
p^2=\det_qw\q,\hskip20pt 
P^2=\det_{q^{n^2}}W\q.          \label{moment}
\ee
 Our work in section 4 with $\U_q(su(2,f))$, 
the dual to $SL_q(2)$ suggests that we view the additional structure due to $\{p_\mu\}$ as in some
way connected to anyonic degrees of freedom. Considering the case of $q=+i$ we find from 
(\ref{v.XIX}), (\ref{v.XXe}) and (\ref{v.XXf}) that
\be
(p_0^2-p_3^2)^2=P_0^2-P_1^2-P_2^2-P_3^2\q,\q\ie\q (p^2)^2=P^2\q.                                   \label{v.XXp}
\ee
Thus the deformed Laplacian is the square root of the undeformed Laplacian. This means that we can
use it to construct an equation which is Dirac-like in the sense that it is a square root 
of the Klein-Gordon equation.
\be
(p^2\pm m^2)\vert\psi\rangle=0\q,                                             \label{v.XXq}  
\ee
where $m^2$ is assumed real. Using (\ref{v.XXp}), and defining $M=m^2$,
we find that for $q=i$
\be
\eqalign{
(p^2\mp m^2)(p^2\pm m^2)\vert\psi\rangle&=((p^2)^2-m^4)\vert\psi\rangle\q\cr
&=(P^2-M^2)\vert0\rangle=0\q,                                                 \label{v.XXr}
}\ee
and thereby recover the Klein-Gordon equation from the undeformed case.
More generally, if $q=\exp{2\pi ir\over {\tilde n}}$, with $n$ prime relative to $r$ we have
\be
p^2=\det_q\omw=p_0^2-\cos{2\pi r\over {\tilde n}}(p_1^2+p_2^2)-p_3^2\q,                       \label{v.XXab}
\ee 
which by (\ref{v.XIX}) is the $n$th root of the Laplacian on undeformed Minkowski space.
Using this we can construct $n$th roots of the Klein-Gordon equation. For even ${\tilde
n}$ these roots are given by
\be
(p^2-q^{2s}m^2)\vert\psi\rangle=0\q,                                         \label{v.XXac}
\ee
for any integer $s$, so that there are $n$ distinct roots. Each of these equations implies the
Klein-Gordon equation because 
\be
(p^2)^n\vert0\rangle=q^{2sn}m^{2n}\vert0\rangle=m^{2n}\vert0\rangle\q,           \label{v.XXad}
\ee
or in the form analogous to (\ref{v.XXr}),       
\be\eqalign{
\prod_{r=s}^{n+s-1}(p^2-q^{2r}m^2)\vert\psi\rangle
&=\prod_{r=1}^{n}(p^2-q^{2r}m^2)\vert\psi\rangle\q\cr
&=((p^2)^n-m^{2n})\vert\psi\rangle\q\cr
&=(P^2-M^2)\vert\psi\rangle\q,                                                   \label{v.XXae}  
}\ee
where we have made use of (\ref{v.XV}), and defined $M=m^n$. For odd $\tilde{n}$ there is an analogous argument.
This time the $n$th roots of the Klein-Gordon equation are given by
\be
(p^2-q^{s}m^2)\vert\psi\rangle=0\q,                                         \label{v.XXaf}
\ee
and we verify that they imply the Klein-Gordon equation by using
\be
(p^2)^n\vert0\rangle=q^{sn}m^{2n}\vert0\rangle=m^{2n}\vert0\rangle
=M^2\vert0\rangle\q.           \label{v.XXag}
\ee      
Equations (\ref{v.XXac}) and (\ref{v.XXaf}) have a clear interpretation as anyonic
Dirac-like equations.\par 

Although (\ref{v.XII}) prevents us from setting $A$ or $D$ equal to zero, there is nothing to
prevent us from defining a restricted, form of $GL_q(2)$ with $B=C=0$. We will refer to this as
simply {\it restricted} $GL_q(2)$. Apart from dropping the $B$ and $C$ parts, 
the only change to the Hopf structure is
that (\ref{v.XI}) becomes
\be
\Delta (A)=A\otimes A\q,\quad \Delta(D)=D\otimes D\q,           \label{v.XXI}
\ee
so that $A$ and $D$ are grouplike. For the sub-case of restricted $SL_q(2)$, we also
have from (\ref{v.IX}) and (\ref{v.XVIII}),
\be
AD=DA=1\q,						      \label{v.XXII}
\ee
so that we can write $D$=$A^{-1}$.
 Using an approach similar to that adopted in section 4 we can 
add generators to restricted $GL_q(2)$ (equally $SL_q(2)$) to obtain {\it extended} 
$GL_q(2)$.
Specifically, we introduce extra elements $\B$ and $\C$, which we endow with the
Hopf algebraic structure of respectively, ${B\over\[n\]!}$ and ${C\over\[n\]!}$ 
in the \ql limit of the generic case. 
Straightforwardly we obtain
\be
\eqalign{
[A,b]_{q^{n}}&=[A,c]_{q^{n}}=[D,b]_{q^{n}}=[D,c]_{q^{n}}=0\q,\cr
[\B,a]_{q^{n}}&=[\B,d]_{q^{n}}=[\C,a]_{q^{n}}=[\C,d]_{q^{n}}=0\q,\cr
[A,d]&=[D,a]=[\B,c]=[\C,b]=0\q,                                         \label{v.XXIII}    
}\ee 
and
\be
\eqalign{
[A,\B]_{q^{n^2}}&=0\q,\quad[A,\C]_{q^{n^2}}=0\q,\quad [A,D]=0\q,\cr
[\B,\C]&=0\q,\quad [\B,D]_{q^{n^2}}=0\q,\quad[\C,D]_{q^{n^2}}=0\q,          \label{v.XXIV}
}\ee
as well as
\be
\varepsilon(A)=\varepsilon(D)=1\q,\quad \varepsilon(\B)=\varepsilon(\C)=0\q,                 \label{v.XXV}
\ee     
and
\be
\eqalign{
S(A)&=(\det_{q^{n^2}}W)^{-1}D\q,\hskip49pt 
S(\B)=-q^{-n^2}(\det_{q^{n^2}}W)^{-1}\B\q,\cr
S(\C)&=q^{n^2}(\det_{q^{n^2}}W)^{-1}\C\q,\quad 
S(D)=(\det_{q^{n^2}}W)^{-1}A\q.                                                \label{v.XXVI} 
}\ee 
The \*-structure of the elements of the extended $*GL_q(2)$ is
\be
*\left(\matrix{A&\B\cr\C&D}\right)=\left(\matrix{A&\C\cr\B&\D}\right)\q.
\label{restar}
\ee   
These are all the same as the corresponding results for $\{A,B,C,D\}$ in
the unrestricted algebra. However, the coproduct is distinct
since for generic $q$
\be
\Delta\left({b^n\over\[n\]!}\right)=
a^n\otimes{b^n\over\[n\]!}+{b^n\over\[n\]!}\otimes d^n
+\sum_{r=1}^{n-1}q^{r^2-nr}{(a\otimes b)^{n-m}(b\otimes d)^m\over
\[r\]!\[n-r\]!}\q.                                                                \label{v.XXVII}
\ee
Taking the \ql limit we obtain the coproduct structure of $\B$, and 
similarly $\C$. The results are
\be
\eqalign{
\Delta (A)&=A\otimes A\q,\quad \Delta(D)=D\otimes D\q,\cr
\Delta(\B)&=A\otimes\B+\B\otimes D+
\sum_{m=1}^{n-1}q^{m^2-nm}{(a\otimes b)^{n-m}(b\otimes d)^m\over\[m\]!\[n-m\]!}\q,\cr
\Delta(\C)&=\C\otimes A+D\otimes\B+
\sum_{m=1}^{n-1}q^{m^2-nm}{(c\otimes a)^{n-m}(d\otimes c)^m\over\[m\]!\[n-m\]!}\q.     \label{v.XXVIII}
}\ee
The cocommutativity of this coproduct can be verified by a lengthy but straightforward calculation. The fact that
the Hopf structure of extended $GL_q(2)$ is well defined is really just a consequence 
of its status as a natural limit of a well defined generic $q$ Hopf algebra, in which 
the retention of all of the generic $q$ structure required the introduction of new elements
$\B$ and $\C$. From (\ref{v.VIab}) we find that the pairings of $\B$ and $\C$ with $\U_q(sl(2))$
are all null except for
\be
\eqalign{
\langle \sp^n,\B\rangle&=q^{n(n-1)}\q,\cr 
\langle \sm^n,\C\rangle&=q^{-n(n-1)}\q.                                 \label{v.XXIX}   
}\ee
Note that this implies that $\spm^n\neq0$, so that extended $SL_q(2)$ is not dual to
$\U_q(sl(2),f)$ or $\U_q(sl(2))$, but rather to a form of 
$\U_q(sl(2))$ which includes the cyclic irreducible representations from \eq{iv.XIII}. 
Thus if we want to maintain the generic $q$ pairing we 
can choose to restrict, then extend, either of $SL_q(2)$ and $\U_q(sl(2))$, 
but not both. Note
that there are intermediate forms with $b^n=0$, $c^n\neq0$ dual to $\sp^n\neq0$, $\sm^n=0$, etc.
\par
Coproducts similar to those for $\B$ and $\C$ appeared in relation to fractional
supersymmetry in \cite{DMAP2,DMAP3,DMAP4,RSD}, suggesting that these elements can be viewed as a 
generalized supersymmetric extension of restricted $GL_q(2)$.\par
 It seems reasonable to expect that other $Fun_q(G)$ algebras
will have analogous properties when their deformation parameters are
roots of unity. It would also be interesting to see if any such
properties are exhibited by braided Hopf algebras such as $BM_q(2)$
\cite{MajI,MajII}.

\vskip20pt
\noindent
{\bf Appendix A}
\vskip20pt

Here we derive (\ref{iv.XXIb}).
First of all note that
\be
\eqalign{
\qlm{1\over\[n\]!}\prod_{k=1-n}^0\[2J_z+k\]&=\qlm{1\over\[n\]!}\prod_{k=1}^n\[2J_z-n+k\]\q\cr
&=\qlm {q^{2J_zn-n^2}q^{n(n-1)\over2}\over (1-q^{-2})^n\[n\]!}\prod_{k=1}^n
(1-q^{-4j_z+2n-2k})\q.                                                \label{aii}   
}\ee
From \cite{DMAP2}  we have the identity
\be
{1\over\[m\]!}\prod_{k=1}^m(1-\sym q^{-2k})=
{1\over\[m\]!}\sum_{k=0}^m{(-\sym)^k q^{k(k+1)}[m]_{q^2}!\over
[m-k]_{q^2}![k]_{q^2}!}\q.                                            \label{aiii}  
\ee
Setting $m=n$ and taking the limit as \ql we find that
\be
\qlm{1\over\[n\]!}\prod_{k=1}^n(1-\sym q^{-2k})=
\qlm{1\over\[n\]!}(1+(-1)^nq^{n(n+1)}\alpha^n)+\sum_{k=1}^{n-1}{(-\sym)^k q^{k(k+1)}
q^{n(n-1)\over2}\over[n-k]_{q^2}![k]_{q^2}!}\q.                       \label{aiv}
\ee
If we now set $\sym=q^{-4\sz+2n}$, then the first term is well defined, and using 
(\ref{aii}) we find that
\be
\eqalign{
\qlm{1\over\[n\]!}\prod_{k=1-n}^0\[2J_z+k\]&=\qlm{q^{2n\sz-n^2}\over\prod_{k=1}^{n-1}(1-q^{-2k})}
\left({1+(-1)^nq^{n(n+1)}q^{-4\sz n+2n^2}\over1-q^{-2n}}\right)\cr 
&+{q^{2n\sz}q^{-{n(n+1)\over2}}\over(1-q^{-2})^n}\sum_{k=1}^{n-1}{(-q^{(1-4\sz+n)})^k\over
\[n-k\]!\[k\]!}\q.      \label{av}
}\ee     
Finally, using the identity $\prod_{k=1}^{n-1}(1-q^{-2k})=n$ \cite{DMAP2}, and (\ref{ai})
this reduces to
\be
\qlm{1\over\[n\]!}\prod_{k=1-n}^0\[2J_z+k\]=
{q^{2n\sz-n^2}\left({2\sz-\frac{3n}{2}-\frac{1}{2}\over n}\right)}
+{q^{2n\sz}q^{-{n(n+1)\over2}}\over(1-q^{-2})^n}\sum_{k=1}^{n-1}{(-q^{(1-4\sz+n)})^k
\over\[n-k\]!\[k\]!}\q,                                                          \label{avi} 
\ee   
in agreement with (\ref{iv.XXIb}).
\vskip30pt\noindent
\begin{Large}
{\bf Acknowledgements}
\end{Large}
\vskip10pt

{ Thanks to Alan Macfarlane, Jos\'e de Azc\'arraga and Carlos P\'erez
Bueno for many helpful discussions and communications. I gratefully
acknowledge the financial support provided by the E.P.S.R.C. and  St
John's College, Cambridge during the preparation
of this paper.}

\end{document}